\documentclass{ifacconf}

\usepackage{graphicx}      
\usepackage{natbib}        
\usepackage{amsmath}       
\usepackage{amssymb}       
\usepackage[utf8]{inputenc}

\usepackage{xspace}
\usepackage{soul}
\usepackage{mathtools}
\usepackage{makecell}
\usepackage{nicefrac}
\usepackage{float}
\usepackage{wrapfig}
\usepackage{color}

\newcommand{\ie}{\unskip, i.\,e.,\xspace}
\newcommand{\eg}{\unskip, e.\,g.,\xspace}

\newcommand{\sut}{\text{s.\,t.\,}}
\newcommand{\wrt}{w.\,r.\,t.\xspace}

\newcommand{\R}{\ensuremath{\mathbb{R}}}

\usepackage{textcomp}

\definecolor{dgreen}{rgb}{0.0, 0.5, 0.0}

\newcommand{\spc}{\ensuremath{\,\,}}	

\makeatletter
\newcommand{\subalign}[1]{%
	\vcenter{%
		\Let@ \restore@math@cr \default@tag
		\baselineskip\fontdimen10 \scriptfont\tw@
		\advance\baselineskip\fontdimen12 \scriptfont\tw@
		\lineskip\thr@@\fontdimen8 \scriptfont\thr@@
		\lineskiplimit\lineskip
		\ialign{\hfil$\m@th\scriptstyle##$&$\m@th\scriptstyle{}##$\crcr
			#1\crcr
		}%
	}
}

\newcommand{\CRIFAC}[2]{©2020 { #1}.\\ This work has been accepted to IFAC for publication under a Creative Commons Licence CC-BY-NC-ND {#2}}

\begin{document}
\begin{frontmatter}

\title{Optimal control of centrifugal spreader} 

\CRIFAC{\author[First]{Franz Rußwurm} 
\author[First]{Pavel Osinenko} 
\author[First]{Stefan Streif}}{DOI: 10.1016/j.ifacol.2020.12.239}

\address[First]{Technische Universität Chemnitz, Automatic Control and System Dynamics Laboratory, 
	Germany (e-mail: \{franz.russwurm;pavel.osinenko;stefan.streif\}@etit.tu-chemnitz.de)}

\begin{abstract}
	Achieving an evenly distributed fertilization spread pattern is a complex technical task.
	A corresponding control algorithm must account for the tractor movement, the settings of the spreader, the prescribed dosage as well as machine constraints.
	It dictates, in particular, the fertilization process needs be estimated ahead to achieve an optimized spread pattern.
	The presented work is concerned with the development of a predictive control scheme for optimized fertilizer application using modeling of the tractor moving on the field and the spread pattern in form of a crescent behind the tractor.
	In particular, the form of the spread pattern is modeled via four normal distributions, two for each side of the pattern.
	The control goal is to achieve a desired fertilizer distribution on the field.
	The study presents three algorithms for comparison: a one-step optimization and two approaches using model-predictive control -- one with a simplified model of the spread pattern in the prediction horizon, and one with a comprehensive one model, respectively.
	The best results are obtained with model-predictive control using the comprehensive model.
\end{abstract}


\end{frontmatter}

\section{Introduction}


	With ever stricter regulations on the amount of the applied fertilizer in agriculture, the demand for optimized fertilization mechanisms increases \citep{Pascale2018-Waterandfertilization,Thompson2015-OptimizingNitrogen}.
	Inefficient fertilizer application leads to unnecessary material waste, harmful environmental effects and reduced crop quantity \citep{Isherwood1998-mineralfertilizer,Sogaard1994-yieldreduction}.
	\citet{Dillon2003-OptimalPath} suggested to optimize the path planning on the field to increase fertization accuracy.
	Such a method cannot be applied if the tramlines are already fixed \eg due to sowing.
	Another methodology is based on the settings of the spreader itself given a fixed tramline.
	It may use \eg a cost function based on the deviation from a prescribed dosage of fertilizer, in the form of a map, which is optimized \citep{Palmer2003-FieldOperations, Virinr2019-optfertilization}.
	This map is obtained \eg by monitoring the state of the plant growth with a sensor attached to the tractor while driving on the field \citep{Sophocleos-PrecisionAgriculture}.
	Such an approach in turn requires a model of the spread pattern, which varies strongly based on the application technique.
	Nowadays centrifugal spreaders with dual spinnning discs are the most used applicators \citep{Liedekerke2008-modelingcentrifugalspreader}.
	The setup in this case can be seen in Fig. \ref{fig:Schematic_representation_of_the_fertilizer_spreader}.
	The fertilizer proceeds from the hopper, a bin attached in the back of the tractor containing the fertilizer granules, onto the discs of the spreader.
	The amount of fertilizer coming from the hopper as well as the revolutions per minute of the discs (RPM) are controlled by the farmer.
	The centrifugal effect ejects the fertilizer granules from the discs resulting in a spread pattern.
	\citet{Fulton2003-Simulation} suggested a simple model based on a rectangular pattern and divided into stripes of equal amount of fertilizer.
	A more accurate model is obtained by simulating and predicting each particle of the fertilizer \citep{Dintwa2004-ParticelFlow,Marinello2017-IntegratedApproach,Villette2008-CentrifugalSpreading} in combination with image processing \citep{Cool2015-ImageBased,Hijazi-SpreadPattern}, although this is associated with high computational costs.
	A model of a twin disc centrifugal spreader more simple to calculate  with a spread pattern in the form of a crescent of two normal distributions was introduced in \citep{Virinr2019-optfertilization,Colin1997-PhD,Olieslagers1997-PhD}.
	\citet{Virinr2019-optfertilization} used this model to optimize the parameters of the spread pattern during the process of fertilization.
	In this work, the model is refined such that the distribution parameters depend on the disc RPM and the mass flow rate, thus the settings of the fertilizer spreader are optimized rather than the parameters of the distribution.
	The only optimization done with this model by \citet{Virinr2019-optfertilization} used an offline approach with a decomposition of the field in subareas consisting of three tramlines and optimizing the parameters of the center tramline.
	So far, however, model-predictive control (MPC) has not yet been used for the centrifugal spreader settings.
	MPC is nowadays a broadly applied and well-established approach of optimal control; it handles nonlinear multiple-input multiple-output systems as well as constraints.
	In this paper, MPC is applied to optimizing the settings of a twin disc centrifugal spreader.

	A precise application of fertilizer ensures a greater profit by not damaging the plants and saving fertilizer, while also complying with regulations.
	Although a precise application is essential, farmers still tend to use only one specific setting per tramline.
	The model introduced in this paper combined with optimal control techniques such as MPC secures a more precise application of fertilizer.
	
%


	The next section states the optimization problem of applying fertilizer on arable land based on a map with prescribed dosages.
	In Section \ref{sec:Model_of_Fertilization}, the model of the tractor and the field along with the spread pattern is introduced.
	Section \ref{sec:Control_algorithm_for_optimal_fertilization} deals with the optimization algorithms.
	For comparison sake, three optimization approaches are used.
	The first one uses optimization based on the current time step only.
	The second and third approaches utilize model-predictive control (MPC) and thus look ahead several time steps.
	The first of those uses a simplified model of the spread pattern throughout the prediction horizon.
	The second utilizes a more comprehensive model (see details in Section \ref{sec:Control_algorithm_for_optimal_fertilization}).
	The last section presents the results.

\section{Problem Statement}\label{sec:Problem_Statement}

\begin{figure}[h]
\begin{center}
\includegraphics[width=6.5cm]{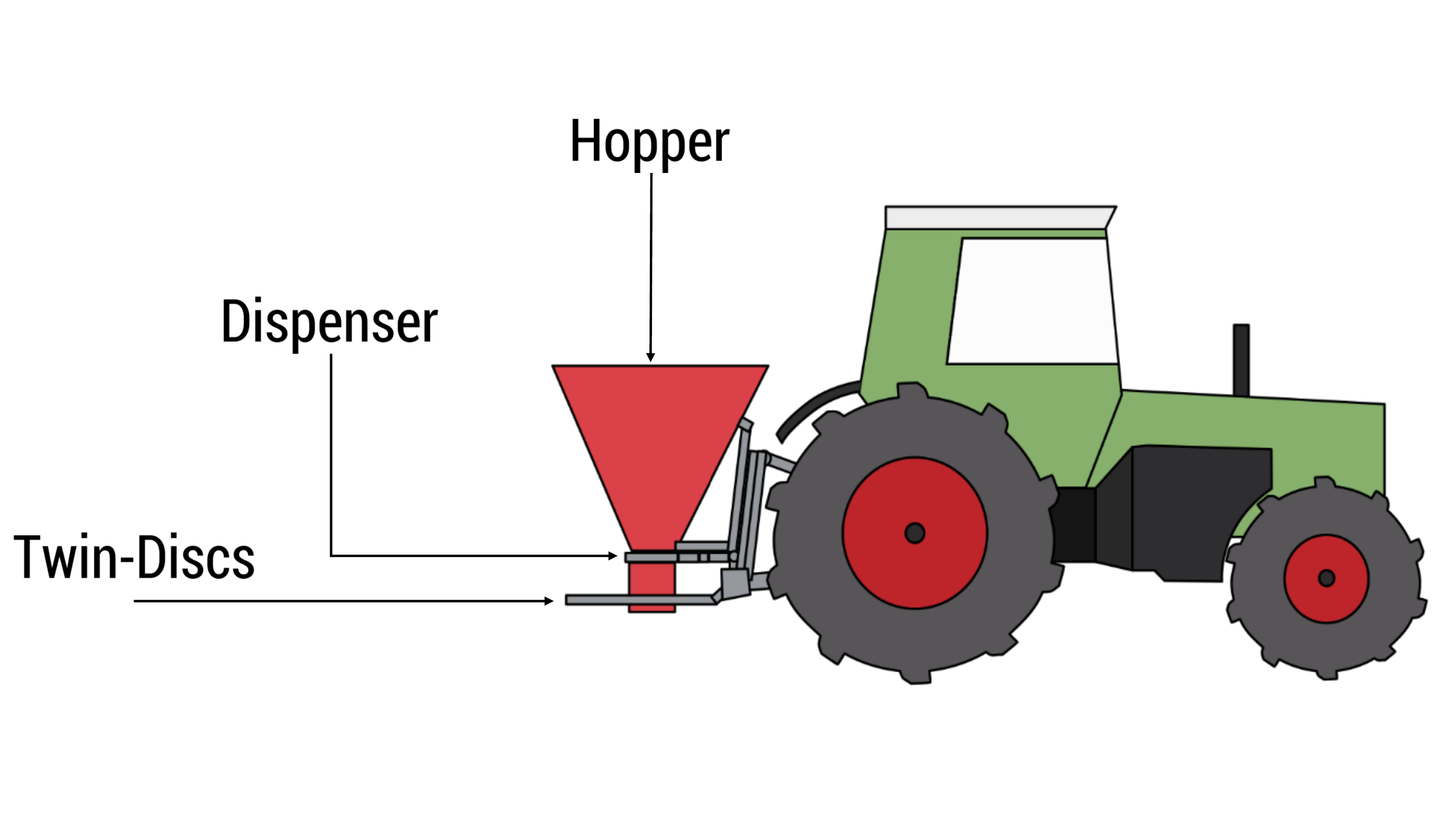}    
\caption{Schematic representation of the centrifugal spreader.} 
\label{fig:Schematic_representation_of_the_fertilizer_spreader}
\end{center}
\end{figure}

	This paper focuses on optimal control of centrifugal spreader implemented on a tractor moving on fixed tramlines.
	The basis for optimization is a field map with a desired amount and distribution of a fertilizer.
	The centrifugal spreader used here is twin-disc with a dispenser regulating the material quantity being poured onto the discs by a hopper, depicted in Fig. \ref{fig:Schematic_representation_of_the_fertilizer_spreader}.
	The discs are driven at independent speeds.
	A crescent-formed spread pattern is assumed here.

	Proceed now to the cost function.
	It is based on the discrepancy between the actual and desired fertilizer applied over a given field.
	As the process goes, the actual fertilization distribution over the field changes.
	The spreader settings are to be controlled subject to process constraints.
	In other words, a problem of the form

	\begin{align}
		\begin{split}
			& \underset{u}{\min} \; J( \mathcal Q (T,u),P) \\
			\\
			\sut \spc & \space \dot{\mathcal Q} (t, u) = f(t,u) \\
			& C_{\text{min}} \leq C(u) \leq C_{\text{max}},
		\end{split}
	\end{align}
	
	where the cost function $J$ penalizes the difference of the actual applied amount of fertilizer $\mathcal Q$ and the prescribed dosage $P$ given by the map.
	The time $T$ denotes the duration of the fertilization process.
	The change of $\mathcal Q$ over time is described by a function $f(t,u)$, where $u$ represents the controls of the centrifugal spreader which are to be optimized \ie the mass flow rate of the fertilizer and the revolutions per minute (RPM) of the discs.
	The restriction on the control $u$ are ensured by a function $C(u)$ whose values are bounded by constants $C_{\text{min}}$ and $C_{\text{max}}$, both related to machine constraints.


\section{Model of fertilization process}\label{sec:Model_of_Fertilization}

	In this section, the model of the fertilization process with a twin-disc spreader used for the development of the control algorithms is described.
	First, the motion of the tractor is modeled via a simple non-holonomic integrator while neglecting traction dynamics, unnecessary for the purposes of the current work.
	In practice, the tractor location is usually fetched from a GPS system.
	Two normal distributions are used to model the spread pattern for each disc.
	One represents the amount of fertilizer applied with respect to the distance from the spreader to the center of the spread pattern, the other features the angle of the spread pattern behind the tractor.
	Both the left and the right patterns are summed up to calculate the resulting dosage of fertilizer applied.
	This model is partly based on the one in \citep{Virinr2019-optfertilization}, where certain parameters of the normal distribution are controlled directly.
	In this work, the distribution parameters depend on the disc RPM and the mass flow rate, thus they are not influenced directly.

\subsection{Kinematic model of the tractor}

	In the fertilization process, the motion of the tractor must be taken into account since, in particular, its speed influences the spread pattern.
First, a simple kinematic model of a wheeled vehicle reads:

\begin{align}\label{eqn:non_holonomic_robot}
\begin{split}
\dot x (t) &= \cos(\varphi(t)) u_1 \\
\dot y (t) &= \sin(\varphi(t)) u_1 \\
\dot \varphi (t) &= u_2.
\end{split}
\end{align}

	Here, the control $u_1 \in \R^2$ represents the speed of the vehicle and $u_2 \in \R$ is the speed of turning.
	The variables $x$ and $y$ denote the location of the tractor on the field while $\varphi$ expresses the direction of driving.

%
%

\subsection{Model of the Field}\label{sec:The_Field}

	To track the amount of fertilizer applied to the field, the latter is divided into squares.
	For simplicity, it is assumed that the number of squares in both $x$ and $y$ directions is the same, call it $N$.
	Let the matrix $A \in \R^{N \times N}$ represent the current amount of fertilizer on the field and the matrix $P \in \R^{N \times N}$ -- the prescribed dosage for each square.
	The row and column indices determine the position of a square on the field, while the matrix entries of $A$ and $P$ themselves are the respective amounts of fertilizer.

\subsection{Model of the spread pattern}



\begin{figure}[h]
\begin{center}
\includegraphics[width=6.5cm]{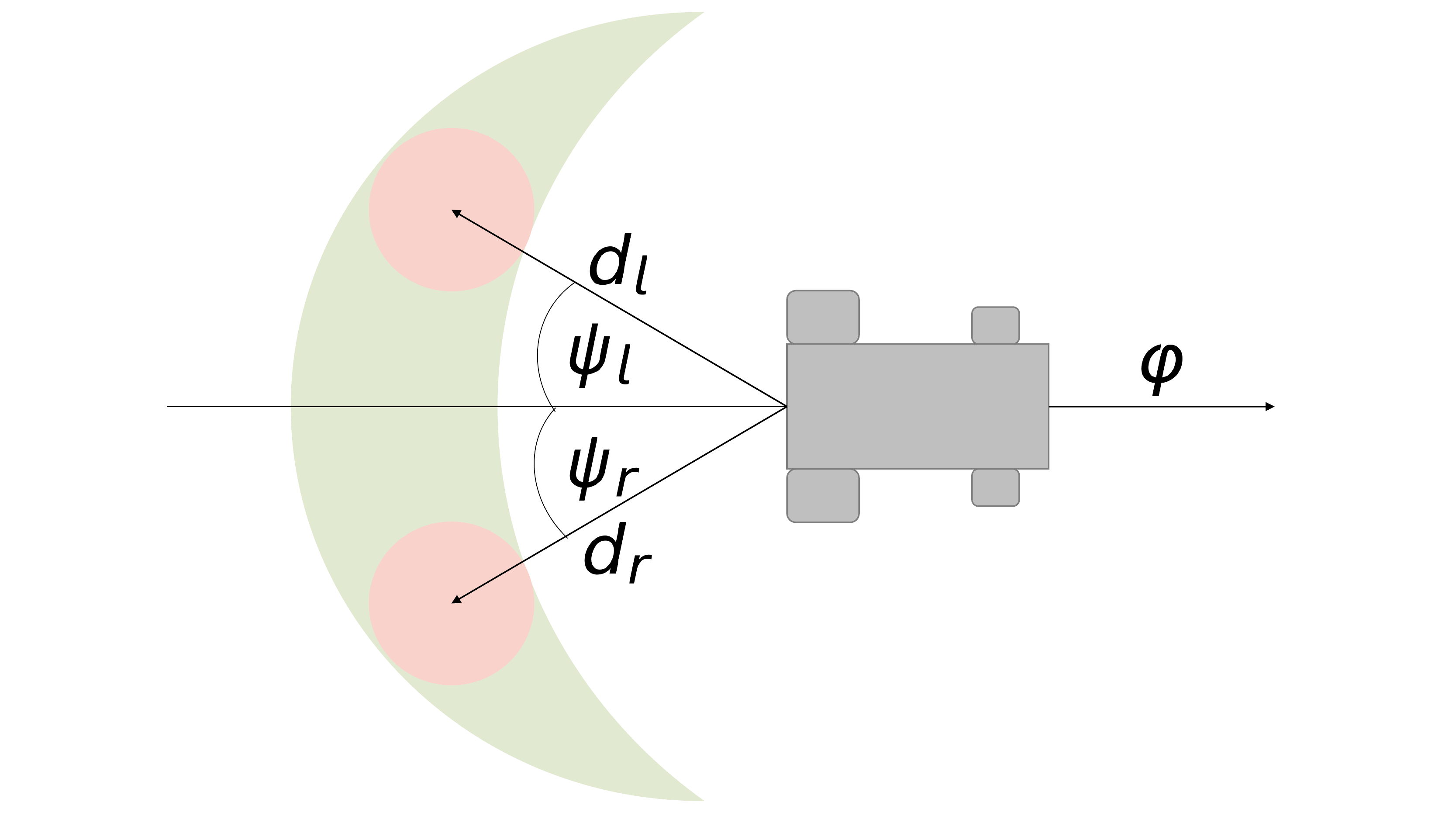}    
\caption{Schematic representation of the tractor.} 
\label{fig:Schematic_representation_of_the_tractor}
\end{center}
\end{figure}

	The spread pattern is assumed here in the form of a crescent as displayed in Fig. \ref{fig:Schematic_representation_of_the_tractor}.
	There are two patterns -- one created by the left disc, one by the right disc.
	They are calculated independently.
	The actual distribution within these spread patterns is modeled via two normal distributions for each side.
	One of these distributions depends on the distance of the center of the spread pattern to the location $\begin{pmatrix} x, & y \end{pmatrix}^T$ of the tractor, the other one depends on the angle between the center of the spread pattern and the opposite direction of driving.
	The combination of these two normal distributions leads to a spread pattern in the form of a crescent.
	The two resulting spread patterns, one for each side, are summed up to give the total distribution.
	The distance between the center of the spread pattern and the location of the tractor is given by $d$.
	The variable $\psi$ represents the angle between the center of the left, respectively the right, side of the spread pattern and the opposite of the direction $\varphi$ of driving.
	An illustration of $d$ and $\psi$ is shown in Fig. \ref{fig:Schematic_representation_of_the_tractor}.
	Both $d$ and $\psi$ are differentiated for the left and the right spread pattern, marked with indices $l$ and $r$ in the following.
	Since the two discs can spin with different speed, the normal distributions for the left and the right disc can have different means and standard deviations.
	The standard deviations for the distance between the spread pattern and the tractor are given by $\sigma_{d_r}$ for the right disc and $\sigma_{d_l}$ for the left.
	The standard deviation \wrt the depicted angles of the spread pattern behind the tractor are described by $\sigma_\psi$, respectively for the right and the left side.

	To calculate the actual distribution at a point $\begin{pmatrix} a, & b\end{pmatrix}^T$ on the field, the difference between the distance of $\begin{pmatrix} a, & b\end{pmatrix}^T$ from the location of the tractor and $d$ as well as the difference of the angle $\psi$ and the angle between $\overrightarrow{\begin{pmatrix} x \\ y \end{pmatrix} \begin{pmatrix} a \\ b\end{pmatrix}}$ and the opposite direction of driving $\varphi + \pi$ need to be determined.
	To compute the distance of the point $\begin{pmatrix} a, & b \end{pmatrix}^T$ to the location of the tractor, the Euclidean norm is used.
	From this value $d$ is substracted. 
	The result of this calculation, named $X$, is the difference of the distance from the center of the spread pattern to the tractor and the distance from $\begin{pmatrix} a, & b \end{pmatrix}^T$ to the tractor which reads as

\begin{align}\label{eqn:Definition_X}
X = \sqrt{(x-a)^2 + (y-b)^2} - d .
\end{align}

	To calculate the difference between the previously described angles, first the angle $\theta$ between the opposite direction of driving $\varphi + \pi$ and $\overrightarrow{\begin{pmatrix} x \\ y \end{pmatrix} \begin{pmatrix} a \\ b\end{pmatrix}}$ must be calculated, which reads as:

\begin{align}
\begin{split}
\theta & = \\
& \arccos \left( \frac{\cos(\varphi + \pi) (a-x) + \sin(\varphi + \pi) (y-b)}{\sqrt{(a-x)^2 + (b-y)^2}} \right).
\end{split}
\end{align}

	Furthermore, turning clockwise or counterclockwise \wrt $\varphi + \pi$ must be verified.
	This condition can be checked with the help of the cross-product $\mathfrak c$, which reads as:

\begin{align}
\mathfrak c = (b-y) \cos(\varphi + \pi) - (a-x) \sin(\varphi + \pi).
\end{align}

If $\mathfrak c < 0$, the angle $\theta$ is changed to $-\theta$.
Finally, the difference $Y$ of $\theta$ and $\psi$ is calculated as:

\begin{align}\label{eqn:Definition_Y}
Y = \theta - \psi .
\end{align}

	To calculate the quantity of fertilizer which is applied at $\begin{pmatrix} a, &\ b \end{pmatrix}^T$, the mass flow rate $D$ is multiplied with the respective values of the normal distribution \wrt the angle and the distance, which reads as

\begin{align}\label{eqn:quantity_left_or_right}
\begin{split}
& q \left( \begin{pmatrix} a \\ b \end{pmatrix}, \begin{pmatrix} x \\ y \end{pmatrix}, D, d, \sigma_d, \psi, \sigma_\psi \right) \\
= \, & \frac{D}{2 \pi \sigma_d \sigma_\psi} \cdot \exp \left( \frac{-X^2}{2 \sigma_d^2} \right) \cdot \exp \left( \frac{-Y^2}{2 \sigma_\psi^2} \right) .
\end{split}
\end{align}

	Since this calculation is done for the left and the right disc separately, the total quantity of fertilizer applied is given by

\begin{align}
\begin{split}
& q_{\text{tot}}\left( \begin{pmatrix} a \\ b \end{pmatrix}, \begin{pmatrix} x \\ y \end{pmatrix}, \begin{pmatrix} D_r \\ D_l \end{pmatrix}, \begin{pmatrix} d_r \\ d_l \end{pmatrix}, \begin{pmatrix} \sigma_{d_r} \\ \sigma_{d_l} \end{pmatrix}, \begin{pmatrix} \psi_r \\ \psi_l \end{pmatrix}, \begin{pmatrix} \sigma_{\psi_r} \\ \sigma_{\psi_l} \end{pmatrix} \right) \\ \\
& = q_r \left( \begin{pmatrix} a \\ b \end{pmatrix}, \begin{pmatrix} x \\ y \end{pmatrix}, D_r, d_r, \sigma_{d_r}, \psi_r, \sigma_{\psi_r} \right) \\
& \, + q_l \left( \begin{pmatrix} a \\ b \end{pmatrix}, \begin{pmatrix} x \\ y \end{pmatrix}, D_l, d_l, \sigma_{d_l}, \psi_l, \sigma_{\psi_l} \right) .
\end{split}
\end{align}

	The variables $D$, $d$, $\sigma_d$, $\psi$ and $\sigma_{\psi}$ all depend on the settings of the centrifugal spreader.
	Therefore, they are time-varing.
	Since these quantities are only calculated for a certain point in time, the function $q_{\text{tot}}$ is integrated over the whole time horizon of the process of applying fertilizer, denoted with $[t_0, T]$, to get the total quantity $\mathcal Q$ of fertilizer applied at the point $\begin{pmatrix} a, & b \end{pmatrix}^T$:

\begin{align}\label{eqn:quantitiy}
\mathcal Q \left( T, \begin{pmatrix} a \\ b \end{pmatrix} \right) = \int_{t_0}^T \, q_{\text{tot}}(t) \, dt
\end{align}


	This calculation is done for each center point $\begin{pmatrix} a, & b\end{pmatrix}^T$ of all the squares on the field.
	The quantity of applied fertilizer is then saved in the matrix $A$.

%
%

\subsection{Disc rotation speed and spread pattern}

	The variables $d$, $\sigma_d$, $\psi$ and $\sigma_\psi$ depend on the disc RPM.
	The dependencies were modeled with the help of spreading charts and the usage of regression. 
	Linear regression was used for the dependency of the distance $d$ of the center of the spread pattern behind the tractor on the disc RPM.
%
%
	The standard deviations $\sigma_d$ and $\sigma_\psi$ as well as the angles $\psi$ were modeled as quadratic functions.
%
%
	It is important to note that for each fertilizer type and for each spreading disc, the spread pattern and the associated variables change.
	Therefore the regression modeling must be performed for each specific machine.

\section{Control algorithm for optimal fertilization}\label{sec:Control_algorithm_for_optimal_fertilization}

	This section is concerned with the design of a model-predictive controller to optimize fertilizer application in accordance with the desired distribution given by matrix $P$.
	For comparison sake, three algorithms are considered here.
	The first algorithm optimizes the settings \ie the RPM of the discs as well as the mass flow rates, in each time step, only using the data available in this time step \ie the current position of the tractor and the already applied amount of fertilizer.
	The second approach is based on MPC.
	First, a simplified model with a triangle distribution for the prediction horizon.
	The second MPC uses the same normal distribution for the prediction as in \eqref{eqn:quantity_left_or_right}.
	The cost function reads

\begin{align}
\begin{split}
J ( A(t), P ) = J(t) := & \, \vert \vert P-A(t) \vert \vert_2^2 \\
=& \, \sum_{i=1}^{N} \, \sum_{j=1}^{N} \, (q_{ij} - a_{ij}(t))^2,
\end{split}
\end{align}

where the variables $q_{ij}$ and $a_{ij}$ mark the entry in the $i$-th row and the $j$-th column of the matrices $P$ and $A$.
	Notice that the applied amount of fertilizer represented in matrix $A$ changes over time throughout the application of the fertilizer.
	The control constraints read

\begin{align}\label{eqn:restrictions_control}
\begin{split}
D_{\text{min}} \leq D \leq D_{\text{max}} \\
\text{RPM}_{\text{min}} \leq \text{RPM} \leq \text{RPM}_{\text{max}},
\end{split}
\end{align}

respectively, for the right and the left side.
	Both constraints for the mass flow rate $D$ and the disc RPM determine the maximal and the minimal values, respectively.
	These restrictions are dictated by safety concerns and machine specifications of the centrifugal spreader.
	To ensure that the control values do not change arbitrarily fast, additional restrictions for the derivation of the controls are introduced:

\begin{align}\label{eqn:restrictions_dcontrol_dt}
\begin{split}
\vert \vert \dot D \vert \vert_2 & \leq \dot D_{\text{max}} \\
\vert \vert \dot{\text{RPM}} \vert \vert_2 & \leq \dot{\text{RPM}}_{\text{max}} ,
\end{split}
\end{align}

	respectively, for the right and the left side.
	Let the input constraint set $\mathcal U$ contain the controls $u = (D_l, D_r, \text{RPM}_l,$ $\text{RPM}_r)$ consisting of pairs of the mass flow rates $D_l$ and $D_r$ as well as the disc $\text{RPM}_r$ and $\text{RPM}_l$ that fulfill the restrictions \eqref{eqn:restrictions_control} and \eqref{eqn:restrictions_dcontrol_dt}.
	The optimal control problem can be stated as:

\begin{align}\label{eqn:optimal_control_problem}
\begin{split}
& \space \underset{u \in \mathcal U}{\min} \; J(T) \\
\\
\sut & \space \dot{\mathcal Q} \left(t, \begin{pmatrix} a \\ b \end{pmatrix} \right) = q_{\text{tot}} \left( t, \begin{pmatrix} a \\ b \end{pmatrix}, \begin{pmatrix} x_0 \\ y_0 \end{pmatrix}, u \right) \\
& \space \begin{pmatrix} a \\ b \end{pmatrix}, \begin{pmatrix} x_0 \\ y_0 \end{pmatrix} \in \mathcal F ,
\end{split}
\end{align}

where $\mathcal F$ denotes the set of all points on the field and $\mathcal U$ is the input constraint set.
	This problem is time-discretized for implementation purposes to yield a cost function for $t_k \in [t_0, T]$ with $k \in \{1, \ldots, n \}$:

\begin{align}
\begin{split}
\hat J (t_k) := \vert \vert P- \hat A (t_k) \vert \vert^2 = \sum_{i=1}^{N} \, \sum_{j=1}^{N} \, (q_{ij} - \hat{a}_{ij}(t_k))^2 ,
\end{split}
\end{align}

where the matrix $\hat A$ is calculated with the discretized version $\hat{\mathcal Q}$ of $\mathcal Q$ from \eqref{eqn:quantitiy}.
	The discretized optimal control problem reads as:

\begin{align}\label{eqn:discrete_problem}
\begin{split}
& \space \underset{u \in \mathcal U}{\min} \; \hat J (T) \\
\\
\sut & \space \hat{\mathcal Q} \left(t_k, \begin{pmatrix} a \\ b \end{pmatrix} \right) = \sum_{i=1}^{k} \, q_{\text{tot}} \left( t_i, \begin{pmatrix} a \\ b \end{pmatrix}, \begin{pmatrix} x_0 \\ y_0 \end{pmatrix}, u \right) \\
& \space \begin{pmatrix} a \\ b \end{pmatrix}, \begin{pmatrix} x_0 \\ y_0 \end{pmatrix} \in \mathcal F \\
& \space k \in \{1, \ldots, n \} .
\end{split}
\end{align}

	As mentioned above, the first approach to find a solution to the optimization problem \eqref{eqn:discrete_problem} is to optimize the control at each time point $t_k$ with only the data available at this time.
	For the second approach, MPC with a simplified model to predict the amount of fertilizer applied during the prediction horizon is used.
	The normal distributions are replaced with triangle distributions.
	This yields for the predicted quantities of applied fertilizer the equation

\begin{align}
\begin{split}
& q\left( t, \begin{pmatrix} a \\ b \end{pmatrix}, \begin{pmatrix} x_0 \\ y_0 \end{pmatrix}, u \right) = \\
& D \left( \frac{-1}{\sqrt{2 \pi} \sigma_d} X + \frac{1}{\sqrt{2 \pi} \sigma_d} \right) \left( \frac{-1}{\sqrt{2 \pi} \sigma_{\psi}} Y + \frac{1}{\sqrt{2 \pi} \sigma_{\psi}} \right)
\end{split}
\end{align}

with $X$ as in \eqref{eqn:Definition_X} and $Y$ from \eqref{eqn:Definition_Y}, for each the left and the right side of the spread pattern.
	This simplified model is used to optimize the control over a prediction horizon of $5$ time steps.
	Then, the control during the first time step is implemented.
	The third and last approach utilizes MPC with the the spread pattern model \eqref{eqn:quantity_left_or_right} also in the prediction horizon.

\section{Results}\label{sec:Results}

	For the example, the tractor drives along three tramlines and turns in between from one tramline into the next.
	This results in driving forward in the first tramline, taking a right turn, driving through the second tramline, taking a left turn and driving through the last tramline, creating an S-shaped pattern.
	While driving through a tramline the control $(u_1, u_2) = (10, 0)$ is applied for a time of $10$ seconds.
	The right turn uses the control $(u_1, u_2) = (4, -\frac{1}{16} \pi)$, while the left turn uses $(u_1, u_2) = (4, \frac{1}{16} \pi)$, both maneuvers take a time of $16$ seconds.
	The duration of the complete trajectory is $68$ seconds, starting in the point $\begin{pmatrix} x_0, & y_0 \end{pmatrix} = \begin{pmatrix} 50, & 100 \end{pmatrix}$.
	
%
%
%

\begin{figure}[h]
\begin{center}
\includegraphics[width=8.4cm]{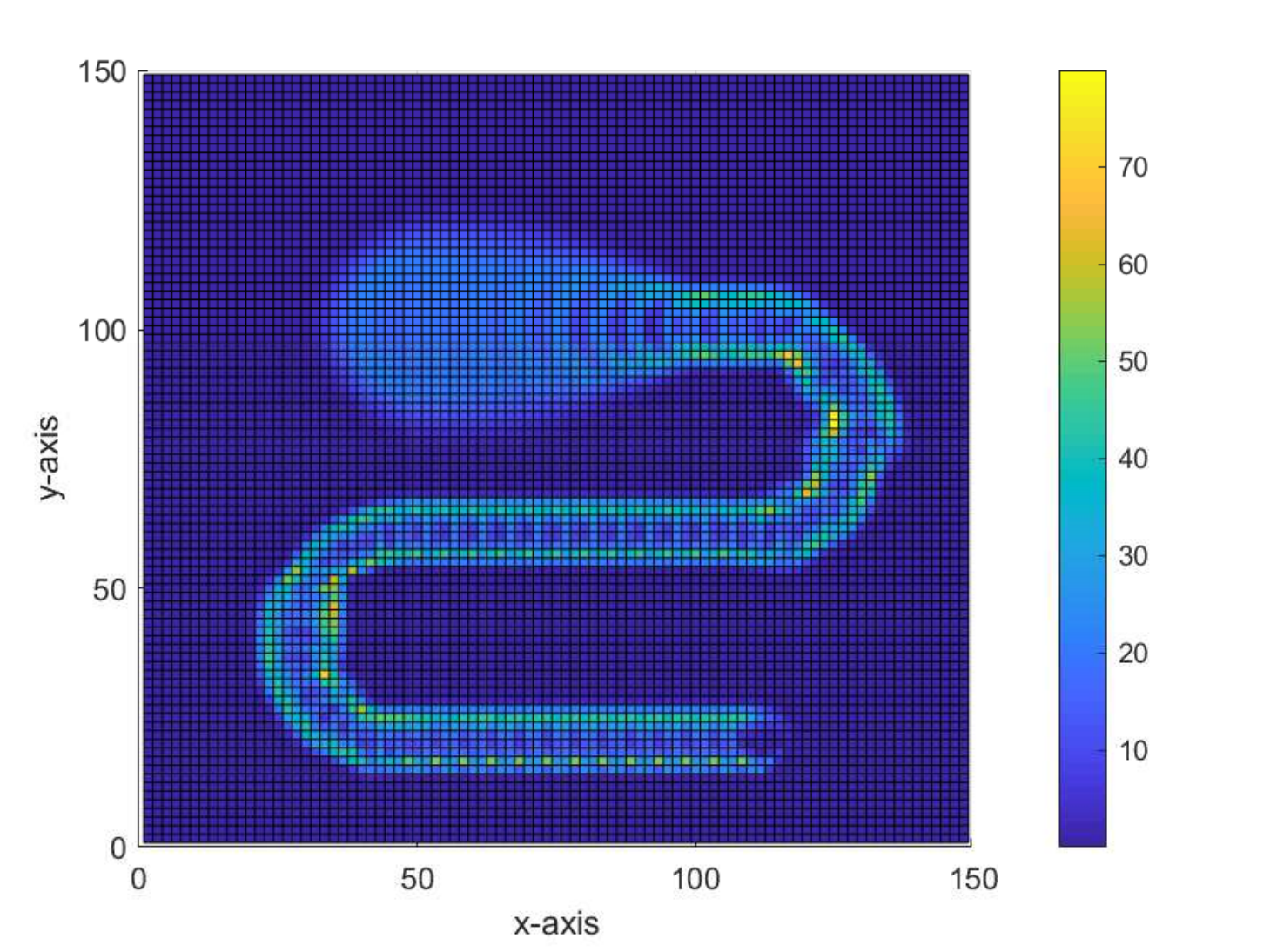}    
\caption{Spread pattern: optimized with fmincon per time step} 
\label{fig:fmincon_spread_pattern}
\end{center}
\end{figure}

	The field size is chosen as $150 \text m \times 150 \text m$.
	It is divided into $90$ segments in each direction, which results in $8100$ squares to measure the amount of fertilizer applied.
	The objective is to achieve an equal prescribed dosage of $20$g of fertilizer for each point on the field.
	The initial values of the mass flow rates for the left and the right side are both $45$g per time step.
	The initial disc RPM is $600$ for both discs.
	These value are choosen arbitrarily.
	The length of the time steps of the discretization is $1$ second.
	The minimal value of the mass flow rates is $0$, while the maximal mass flow rate is set to $200$.
	The minimal admissible RPM is $300$ to avoid a too narrow spread pattern, which would result in high measurement errors of the applied amount of fertilizer, especially if the spread pattern gets smaller than the squares partitioning the field.
	The maximal RPM is $900$, by the machine specifications.
	Furthermore the maximal change of the mass flow rates per time step is limited to $20$g and the maximal change of disc RPM is limited to $100$.
	The first optimization approach with optimization in each time step with only the information available in this time step results in a final spread pattern on the field shown in Fig. \ref{fig:fmincon_spread_pattern}.

\begin{figure}[h]
\begin{center}
\includegraphics[width=8.4cm]{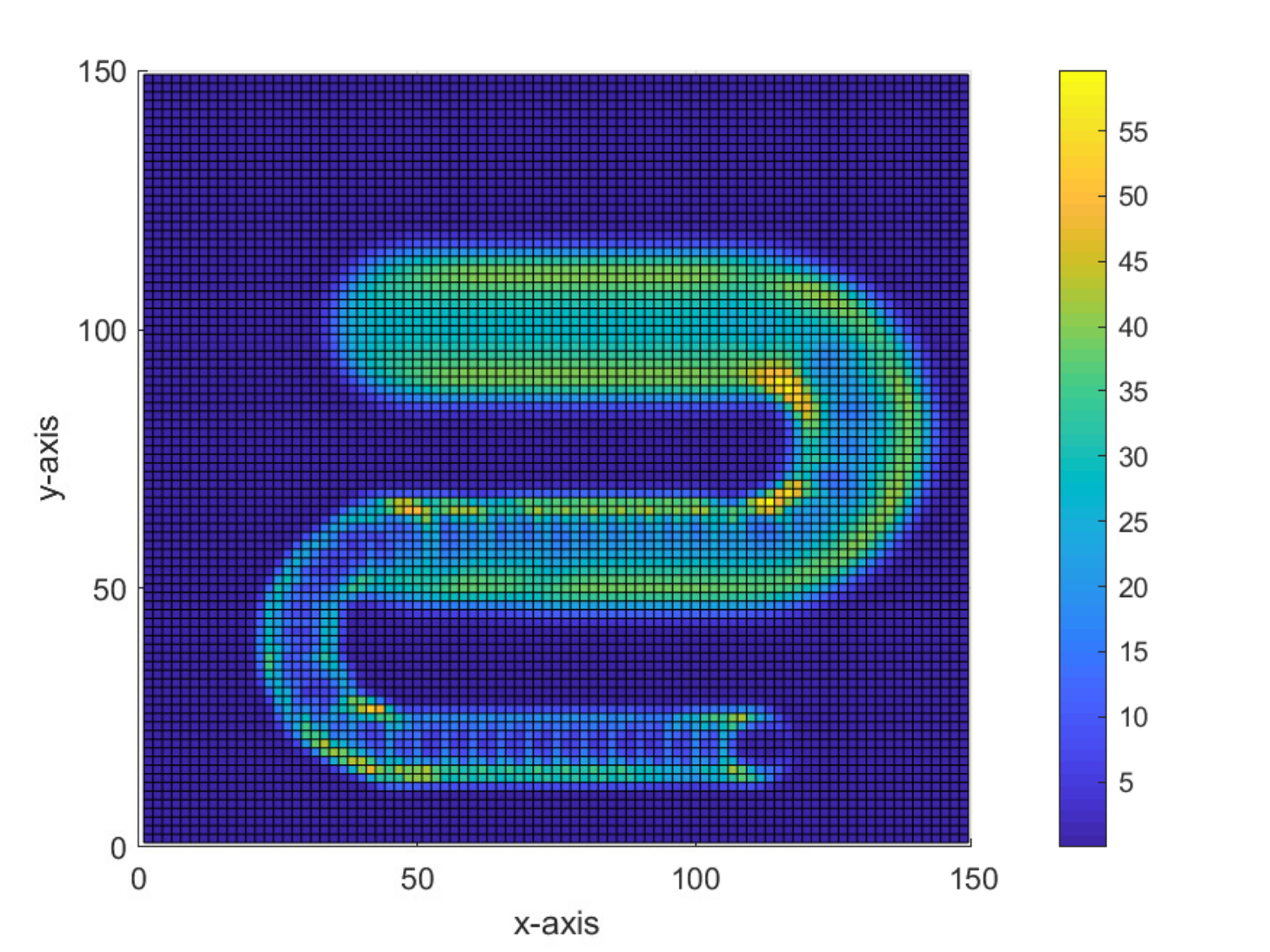}    
\caption{Spread pattern: optimized with MPC, simplified model} 
\label{fig:mpc_simple_spread_pattern}
\end{center}
\end{figure}

	As it can be seen, in the first few time steps a uniform distribution in close range to the desired spread pattern is reached.
	After these first few time steps, the solution presumably gets into a local minimum.
	The disc RPM drop to the minimal value and the spread pattern becomes narrow, with very high maximal values in the center of the left and right spread pattern.
	Although in the center of the total spread pattern the desired value of $20$g is achieved.
	A better spread pattern is obtained by MPC with the simplified model during the prediction horizon of $5$ seconds.
	The resulting spread pattern on the field is shown in Fig \ref{fig:mpc_simple_spread_pattern}.

\begin{figure}[h]
\begin{center}
\includegraphics[width=8.4cm]{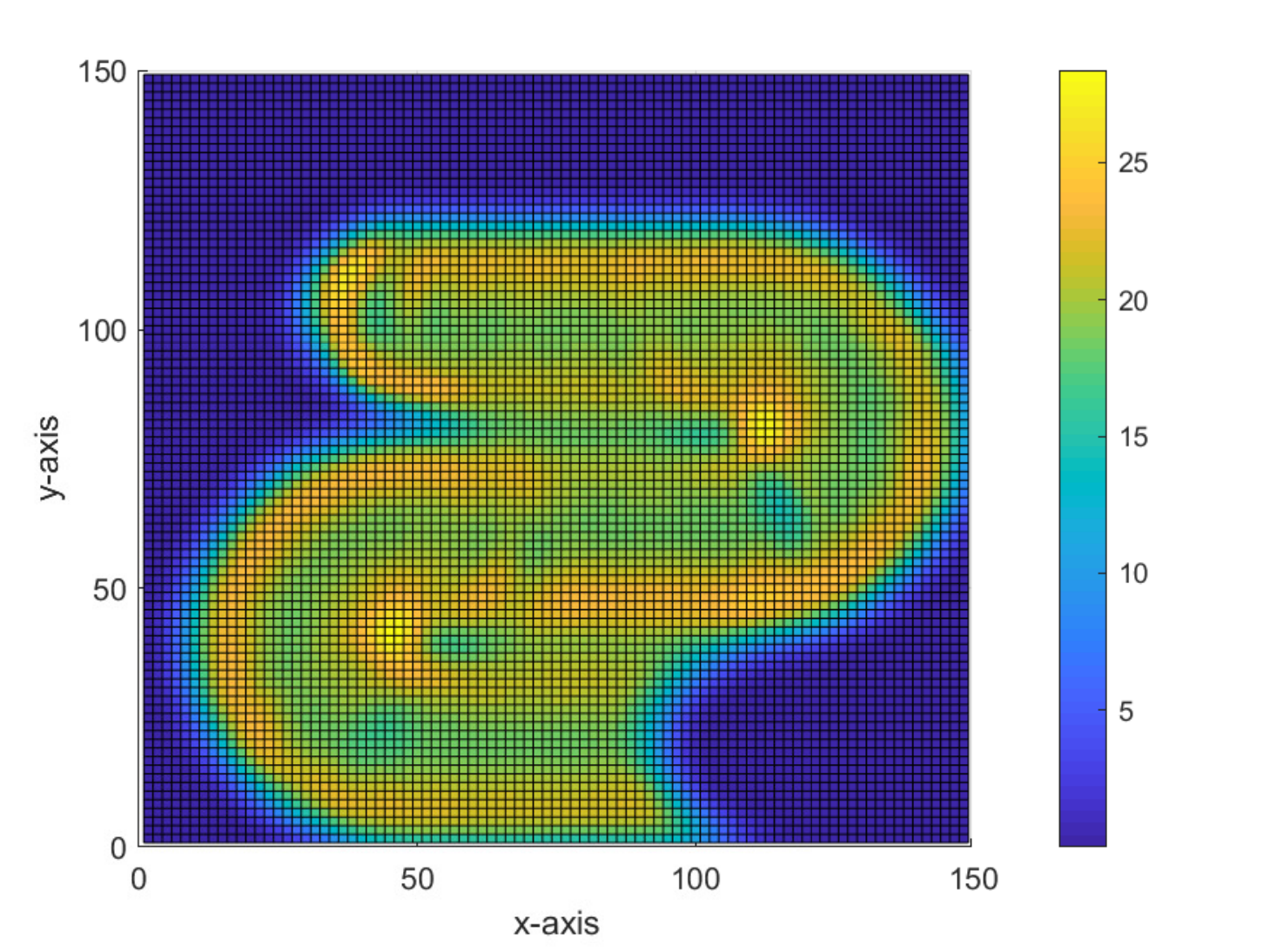}    
\caption{Spread pattern: optimized with MPC, comprehensive model} 
\label{fig:mpc_spread_pattern}
\end{center}
\end{figure}

	The optimal solution in the first tramline results in a spread pattern which achieves the desired distribution with good accuracy.
	After the first turn, the optimal solution gets worse.
	After the second turn, the optimal solution approaches the same local minimum as the solution obtained by using the first method.
	It can be deduced that the disc RPM are getting lower after each turn since the spread pattern gets narrower each time.
	The best spread pattern is achieved by the utilization of MPC with the comprehensive model.
	The results are shown in Fig. \ref{fig:mpc_spread_pattern}.

\begin{figure}[h]
\begin{center}
\includegraphics[width=8.4cm]{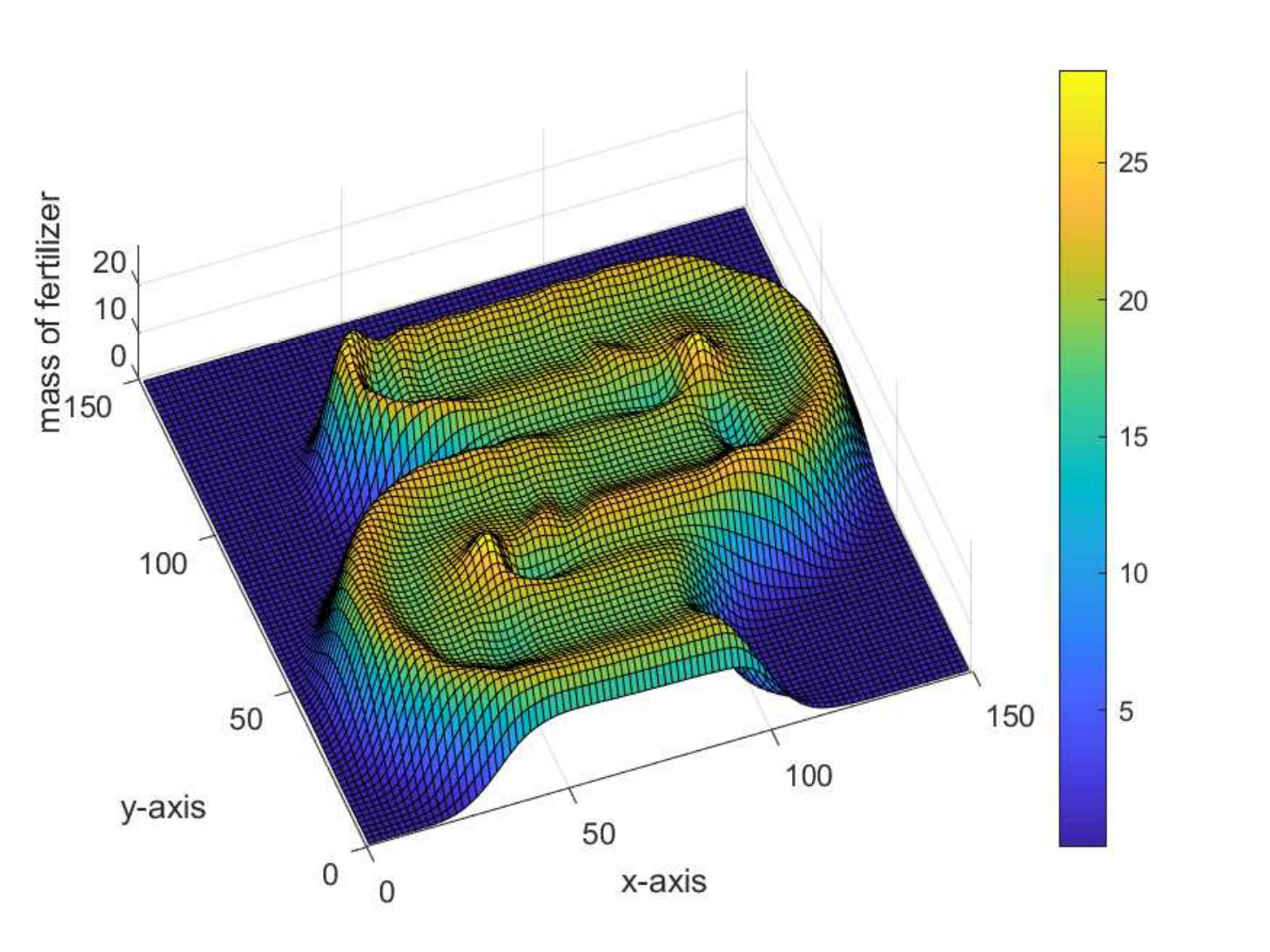}    
\caption{Spread pattern: optimized with MPC, 3D view} 
\label{fig:mpc_spread_pattern_side_view}
\end{center}
\end{figure}

	The three-dimensional view of the spread pattern can be seen in figure \ref{fig:mpc_spread_pattern_side_view}.
	The desired spread pattern is achieved throughout the whole application process.
	Even while driving curves the amount of fertilizer on the field exceeds the prescribed dose by far less than during the usage of the other optimization approaches.
%
%
%
%
	It can be seen that the RPMs clearly react to turns.
	However, unlike in the other optimization approaches, the RPM go back to the maximal value of $900$ to achieve a spread pattern as wide as possible to apply fertilizer at the most parts of the field.
	The computation cost of the MPC approach using the triangle distribution in the prediction horizon is at around $95\%$ of the MPC using the normal distribution.
	This small time saving is due to the fact that most of the computational costs originate from the optimization of the controls.
	Since the simple model contains the same number of optimization variables as the comprehensive model, it only yields a slight reduction of the computational costs in calculation of the spread pattern.
	The first optimization method requires only around $2\%$ of the time needed for MPC approaches, although the resulting solution is the worst.

\section{Conclusion}\label{sec:Conclusion}

	In this paper a model for spread pattern, based on \cite{Virinr2019-optfertilization}, was refined with dependencies of the variables on the controls \ie the mass flow rate and the disc RPM.
	Furthermore three optimization approaches were suggested and compared for the spread pattern optimization, one with optimization for considering only the data in each time step and two using model-predictive control.
	The optimization in each time step without a prediction horizon results in a local minimum of the optimization problem, where in the center part of the spread pattern the demanded amount of fertilizer is achieved, while on the left and right hand side of the spread pattern a too high amount of fertilizer is applied.
	This would lead to harmful over-fertilization effects.
	The first optimization approach with MPC using a triangle distribution for the prediction horizon achieves a better performance.
	Although when taking turns the spread pattern gets worse and approaches the solution of the first optimization attempt.
	This approach only seems to be considerable if one aims for optimization driving along tramlines without taking turns.
	Lastly, an optimization with MPC using a comprehensive model for the prediction horizon was done.
	This approach results in the best solution, the local minimum which attracted the solutions in the first two attempts is completely avoided and the prescribed quantity of fertilizer is closely achieved.
	Using MPC in a real process, one can expect that the
prediction model is not equal to the real plant.
	Therefore, the impact of deviations in the parameters of the model in the prediction horizon to the application model can be tested and analyzed for future work with \eg a robust version of MPC.

\begin{ack}
The project underlying this paper was funded by the Federal Ministry of Education and Research under the grant number 281B300116. The responsibility for the content of this publication lies with the author.
\end{ack}

\bibliography{bib/fertilization}

\begin{thebibliography}{17}
\providecommand{\natexlab}[1]{#1}
\providecommand{\url}[1]{\texttt{#1}}
\providecommand{\urlprefix}{URL }
\expandafter\ifx\csname urlstyle\endcsname\relax
  \providecommand{\doi}[1]{doi:\discretionary{}{}{}#1}\else
  \providecommand{\doi}{doi:\discretionary{}{}{}\begingroup
  \urlstyle{rm}\Url}\fi

\bibitem[{Colin(1997)}]{Colin1997-PhD}
Colin, A. (1997).
\newblock Etude du procede d’epandage centrifuge d’engrais mineraux.
\newblock \emph{PhD thesis}.

\bibitem[{Cool et~al.(2015)Cool, Pieters, Mertens, Nuyttens, Hijazi, Dubois,
  Cointault, and Vangeyte}]{Cool2015-ImageBased}
Cool, S., Pieters, J., Mertens, K., Nuyttens, D., Hijazi, B., Dubois, J.,
  Cointault, F., and Vangeyte, J. (2015).
\newblock Image based techniques for determining spread patterns of centrifugal
  fertilizer spreaders.
\newblock \emph{Agriculture and Agricultural Science Procedia}, 7, 59--63.

\bibitem[{De~Pascale et~al.(2018)De~Pascale, Rouphael, Gallardo, and
  Thompson}]{Pascale2018-Waterandfertilization}
De~Pascale, S., Rouphael, Y., Gallardo, M., and Thompson, R. (2018).
\newblock Water and fertilization management of vegetables: State of art and
  future challenges.
\newblock \emph{European Journal of Horticultural Science}, 83, 306--318.
\newblock \doi{10.17660/eJHS.2018/83.5.4}.

\bibitem[{Dillon et~al.(2003)Dillon, Fulton, Shearer, and
  Kanakasabai}]{Dillon2003-OptimalPath}
Dillon, C., Fulton, J.P., Shearer, S.A., and Kanakasabai, M. (2003).
\newblock Optimal path nutrient application using variable rate technology.
\newblock \emph{Proc. of the Fourth European Conference on Precision
  Agriculture}, 171--176.

\bibitem[{Dintwa et~al.(2004)Dintwa, Liedekerke, Olieslagers, Tijskens, and
  Ramon}]{Dintwa2004-ParticelFlow}
Dintwa, E., Liedekerke, P., Olieslagers, R., Tijskens, E., and Ramon, H.
  (2004).
\newblock Model for simulation of particle flow on a centrifugal fertiliser
  spreader.
\newblock \emph{Biosystems Engineering}, 87, 407--415.

\bibitem[{Fulton et~al.(2003)Fulton, Shearer, Anderson, Burks, Stombaugh, and
  Higgins}]{Fulton2003-Simulation}
Fulton, J.P., Shearer, S.A., Anderson, M.E., Burks, T.F., Stombaugh, T.S., and
  Higgins, S.F. (2003).
\newblock Distribution pattern variability of granular vrt applicators.
\newblock \emph{Transactions of the ASAE}, 46(5), 1311--1321.

\bibitem[{Hijazi et~al.(2014)Hijazi, Vangeyte, Cool, Mertens, Nuyttens, Dubois,
  Cointault, and Pieters}]{Hijazi-SpreadPattern}
Hijazi, B., Vangeyte, J., Cool, S., Mertens, K., Nuyttens, D., Dubois, J.,
  Cointault, F., and Pieters, J. (2014).
\newblock \emph{Predicting spread patterns of centrifugal fertiliser
  spreaders}.

\bibitem[{Isherwood(1998)}]{Isherwood1998-mineralfertilizer}
Isherwood, K.F. (1998).
\newblock Mineral fertilizer use and the environment.
\newblock \emph{IFA}.

\bibitem[{Liedekerke et~al.(2008)Liedekerke, Piron, Vangeyte, Villette, Ramon,
  and Tijskens}]{Liedekerke2008-modelingcentrifugalspreader}
Liedekerke, P., Piron, E., Vangeyte, J., Villette, S., Ramon, H., and Tijskens,
  E. (2008).
\newblock Recent results of experimentation and dem modeling of centrifugal
  fertilizer spreading.
\newblock \emph{Granular Matter}, 10, 247--255.

\bibitem[{Marinello et~al.(2017)Marinello, Pezzuolo, Gasparini, and
  Sartori}]{Marinello2017-IntegratedApproach}
Marinello, F., Pezzuolo, A., Gasparini, F., and Sartori, L. (2017).
\newblock Integrated approach for prediction of centrifugal fertilizer spread
  patterns.
\newblock \emph{Agricultural Engineering International : The CIGR e-journal},
  19, 219.

\bibitem[{Olieslagers(1997)}]{Olieslagers1997-PhD}
Olieslagers, R. (1997).
\newblock Fertilizer distribution modelling for centrifugal spreader design.
\newblock \emph{PhD thesis}.

\bibitem[{Palmer et~al.(2003)Palmer, Wild, and
  Runtz}]{Palmer2003-FieldOperations}
Palmer, R., Wild, D., and Runtz, K. (2003).
\newblock Improving the efficiency of field operations.
\newblock \emph{Biosystems Engineering - BIOSYST ENG}, 84, 283--288.

\bibitem[{S{\o}gaard and Kierkegaard(1994)}]{Sogaard1994-yieldreduction}
S{\o}gaard, H. and Kierkegaard, P. (1994).
\newblock Yield reduction resulting from uneven fertilizer distribution.
\newblock \emph{Transactions of the ASAE, Vol.}, 37(6), 1749--1752.

\bibitem[{Sophocleous and Georgiou(2017)}]{Sophocleos-PrecisionAgriculture}
Sophocleous, M. and Georgiou, J. (2017).
\newblock Precision agriculture: Challenges in sensors and electronics for
  real-time soil and plant monitoring.
\newblock 1--4.

\bibitem[{Thompson et~al.(2015)Thompson, Gallardo, and
  Voogt}]{Thompson2015-OptimizingNitrogen}
Thompson, R., Gallardo, M., and Voogt, W. (2015).
\newblock Optimizing nitrogen and water inputs for greenhouse vegetable
  production.
\newblock \emph{Acta Horticulturae}, 1107, 15--30.
\newblock \doi{10.17660/ActaHortic.2015.1107.2}.

\bibitem[{Villette et~al.(2008)Villette, Piron, Cointault, and
  Chopinet}]{Villette2008-CentrifugalSpreading}
Villette, S., Piron, E., Cointault, F., and Chopinet, B. (2008).
\newblock Centrifugal spreading of fertiliser: Deducing three-dimensional
  velocities from horizontal outlet angles using computer vision.
\newblock \emph{Biosystems Engineering}, 99, 496--507.
\newblock \doi{10.1016/j.biosystemseng.2007.12.001}.

\bibitem[{Virin et~al.(2008)Virin, Koko, Piron, Martinet, and
  Berducat}]{Virinr2019-optfertilization}
Virin, T., Koko, J., Piron, E., Martinet, P., and Berducat, M. (2008).
\newblock Optimization-based approach for a better centrifugal spreading.
\newblock \emph{International Journal of Systems Science}, 39(9), 913--924.

\end{thebibliography}


\end{document}